\documentstyle[editedvolume]{crckapb} 

\begin{opening}
\title{Metallicities of nearby K- dwarfs}

\author{Eira Kotoneva}
\institute{Tuorla Observatory, Piikki\"o, Finland}
\author{Chris Flynn}
\institute{Tuorla Observatory, Piikki\"o, Finland}

\end{opening}

\runningtitle{THE CRCKAPB STYLE FILE}

\begin{document}

One of the main issues in the chemical evolution of galaxies is the so
called ``G-dwarf problem'' (Pagel and Patchett, 1975) in which the
observed stellar metal distribution differs from the predicted in the
sense that there are too many metal deficient stars. The same problem
 was now been seen among K-dwarfs, indicating the problem is a
general one (Flynn and Morell, 1997).

We have used the Hipparcos catalogue to choose all K-dwarfs with
absolute magnitude $M_V$ between $5.5 < M_V < 7.3$, within a radius of
54 parsecs with an apparent visual magnitude $V < 8.2 + 1.1\,{\mathrm
sin}|b|$, where $b$ is the galactic latitude. This sample consists of
642 stars and is complete. 408 of these stars were found from the
Geneva photometric catalogue (Rufener, 1989) and 364 from the
Str\"omgren catalogue of Hauck and Mermilliod (1998).

We have measured metallicities for 248 of these stars using $R - I$
photometry from Bessell (1990) and Geneva and Str{\"o}mgren based
metallicity indicators. For stars with Geneva colours metallicities
were found using existing relations for K-dwrafs (Flynn and Morell,
1997). We have developed a Str\"omgren photometric metallicity
indicator for the stars with Str\"omgren colours. We used 34 G and K
dwarfs from Flynn and Morell (1997) to define a relation between the
Str\"omgren colours, the abundance [Fe/H] and effective temperature
$T_{\rm eff}$. For these stars accurate, spectroscopically determined
metallicity abundances were known and the effective temperatures were
estimated using Cousins $R - I$ photometry from the Gliese catalogue
(Bessell, 1990).

In Fig. 1a the relation between spectroscopically determined
metallicity and the Str\"omgren based metallicity is shown. The scatter
around the one-to-one relation is $\approx 0.2$ dex.

\begin{figure*}[htbp]
\vspace{5.5cm}
\includegraphics{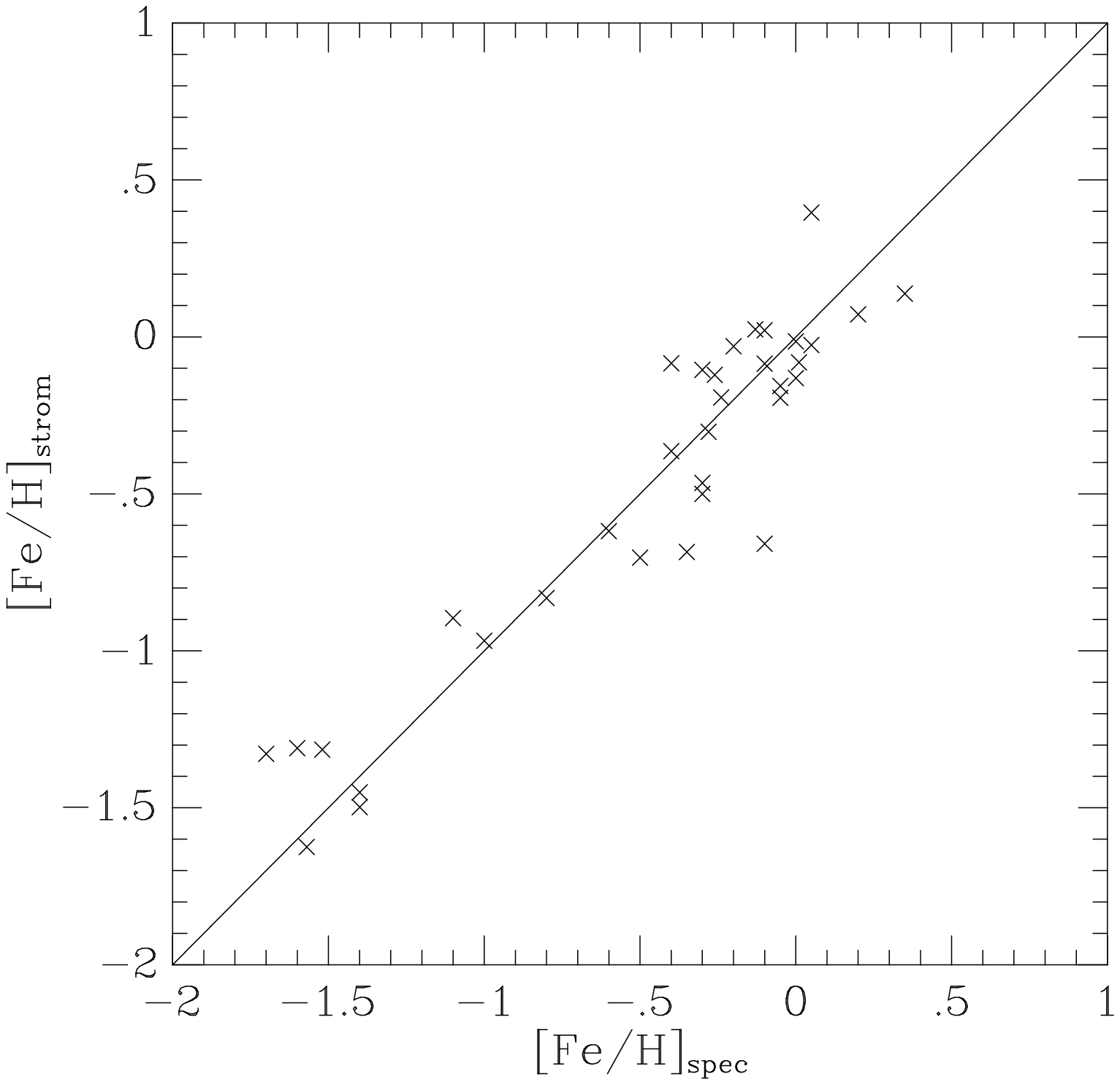}
\caption{Left panel shows the relation between the abundances using
Str{\"o}mgren $b_1$ colour, Cousins $R-I$ and spectroscopically
determined abundances. The right panel shows the histogram of oxygen
abundances of 248 stars. This histogram shows clearly that there is a
``K-dwarf problem''. The Pagel (1989) inflow model is shown as a
dashed line and fits the data quite well.}

\includegraphics{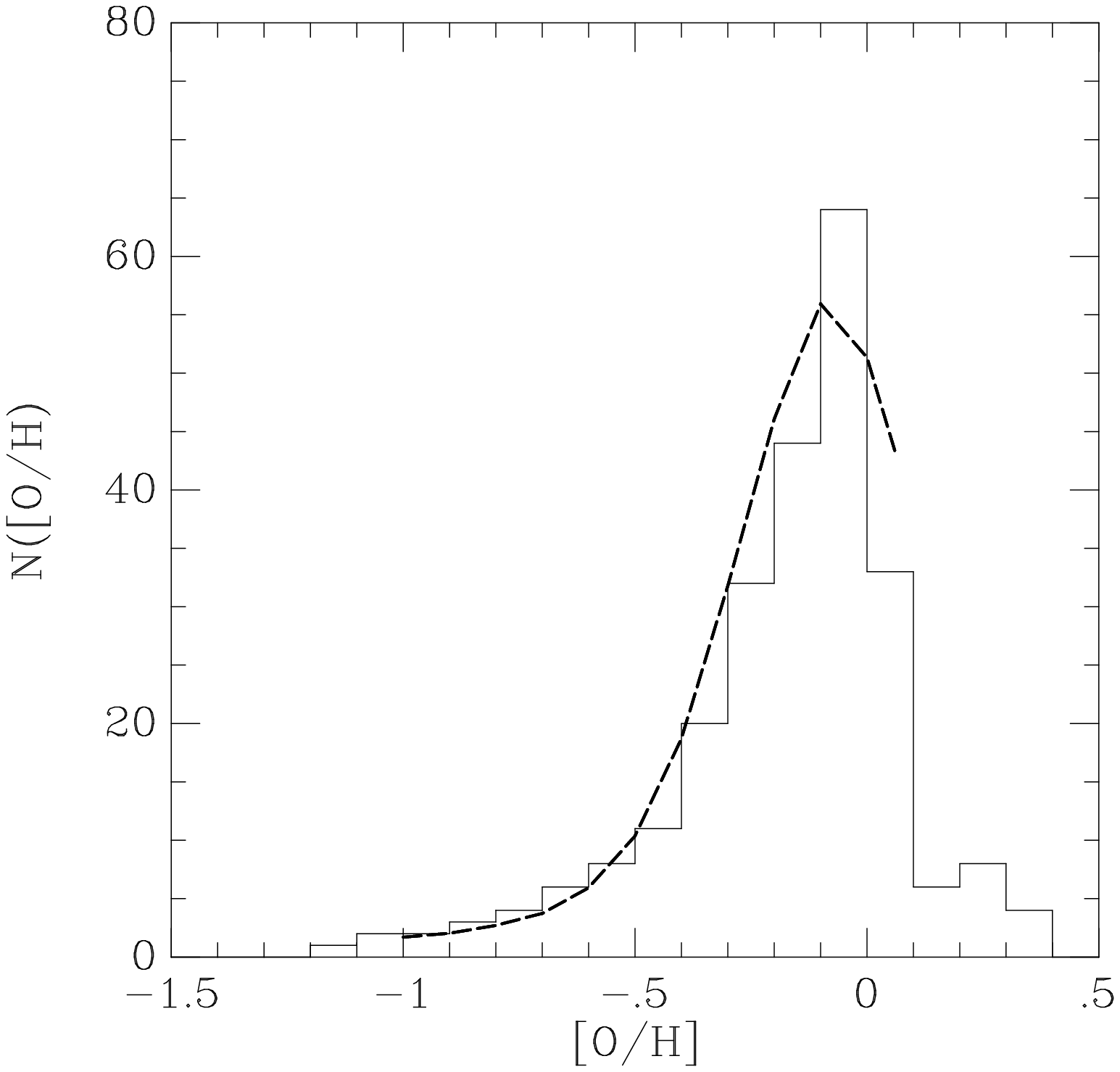}
\end{figure*}

Iron abundances [Fe/H] have been converted to oxygen abundances [O/H]
as in Flynn and Morell (1997) using their equation 6. When comparing
data and models oxygen abundances are more appropriate because oxygen
is produced in short lived stars and can be treated using the
convenient approximation of instantaneous recycling (Pagel, 1989).

The metallicity histogram for all 248 stars is shown in Fig. 1b. In
this figure the K-dwarf problem is clearly seen. The histogram is
peaked near solar metallicity and the number of metal deficient stars
is small.

Note that the sample has not been corrected for two effects. Firstly,
there is a kinematic bias against higher velocity stars in nearby
samples (Sommer-Larsen, 1991). Secondly, the sample is slightly biased
toward too many metal weak stars since these are more likely to be in
the photometric catalogues we used. In the future we plan to observe
the complete sample and thus remove these biases.

\vskip 0.5 truecm
\noindent{\bf References}
\vskip 0.1 truecm

\noindent Bessell, M., 1990, A\&AS, 83, 357

\noindent Flynn, C. and Morell, O., 1997, MNRAS, 286, 617

\noindent Hauck, B., Mermilliod, M., 1998, A\&AS, 129, 431

\noindent Sommer-Larsen, J., 1991, MNRAS, 249, 368

\noindent Pagel, B.E.J., 1989, Rev. Mex. Astr. Astrofis., 18, 153

\noindent Pagel, B.E.J., Patchett, B.E., 1975. MNRAS, 172, 13

\noindent Rufener, F., A\&AS, 78, 469

\end{document}